 \documentclass[10pt]{IEEEtran}
\usepackage[a4paper]{geometry}
\IEEEoverridecommandlockouts

\usepackage{cite}
\usepackage{amsmath,amssymb,amsfonts}
\usepackage{graphicx}
\usepackage{textcomp}
\usepackage{xcolor}
\usepackage{float}
\usepackage{array}
\usepackage{algorithm}
\usepackage{algpseudocode}
\usepackage{bm}
\usepackage{tabularx}
\usepackage{multirow}
\usepackage{booktabs}
\usepackage{stfloats}
\usepackage{ntheorem}

\usepackage{geometry}
\usepackage{subfigure}
\usepackage{stfloats}
\makeatletter
\renewcommand{\fnum@algorithm}{}

\usepackage{epsfig}
\usepackage{epstopdf}

\begin{document}
\title{
	Partial Superimposed Pilot-Aided Sparse Vector Transmission for High-Mobility URLLC}
\author{{Yanfeng~Zhang,~\IEEEmembership{{Member,~IEEE}},}
Jinkai Zheng,~\IEEEmembership{{Member,~IEEE}},~Hui Liang, Wen Wu,~\IEEEmembership{{Senior Member,~IEEE}\vspace{-0pt}},\\ ~Tom H. Luan,~\IEEEmembership{{Fellow,~IEEE}}, Di Zhang,~\IEEEmembership{{Senior Member,~IEEE}}, and Xu~Zhu,~\IEEEmembership{{Senior Member,~IEEE}}
\vspace{-0pt}
\thanks{This work was supported in part by the National Natural Science Foundation of China under Grant 62601166; in part by the Guangdong Basic
and Applied Basic Research Foundation under Grant 2024A1515110033 and 2024A1515110036; in part by the Dongguan Strategic Scientist Teams Project under Grant 20231900700022; in part by the Key Program of the National Natural Science Foundation of China under Grant W2611084; in part by the Guangdong International Science and Technology Cooperation Program under Grant 2025A0505020024; in part by the Key-Area Research and Development Program of Guangdong Province under Grant 2026B0909060001; in part by the Guangdong Provincial Key Laboratory of Space-Aerial Networking and Intelligent Sensing under grant 2024KSYS023. (Corresponding author: Yanfeng Zhang)}
\thanks{Yanfeng Zhang, Jinkai Zheng, and Hui Liang are with the School of Electrical Engineering and Intelligentization, Dongguan University of Technology, Dongguan, China. (e-mails: yfzhang@ieee.org; jkzheng@ieee.org; huiliang@ieee.org).}
\thanks{Wen Wu is with the Department of Strategic and Advanced Interdisciplinary Research, Pengcheng Laboratory, Shenzhen 518000, China. (e-mail: wuw02@pcl.ac.cn).}
\thanks{Tom H. Luan is with the School of Cyber Science and Engineering, Xi’an Jiaotong University, Xi’an, China. (e-mail: tom.luan@xjtu.edu.cn).}
\thanks{Di Zhang is with the School of Intelligent Systems Engineering, Sun Yat sen University, Shenzhen 518107, China (E-mail: dr.di.zhang@ieee.org).}
\thanks{Xu Zhu is with the School of Information Science and Technology, Harbin Institute of Technology, Shenzhen 518055, China. (e-mail: xuzhu@ieee.org).}
\vspace{-0pt}
}

\maketitle

\begin{abstract}
A partial superimposed pilot-aided sparse vector transmission (PSP-SVT) scheme is proposed for short-packet ultra-reliable and low-latency communications in high-mobility scenarios. Unlike conventional full superimposed pilot-aided SVT schemes, the proposed PSP-SVT scheme deploys only a few pilots over a subset of subcarriers. This sparse pilot structure is sufficient for basis expansion model based channel tracking while effectively reducing pilot–data interference. Based on the PSP pattern, an iterative receiver is developed to jointly perform channel estimation and data decoding. The reduced pilot interference in PSP-SVT provides more accurate initial channel estimation and data detection, thereby improving the subsequent iterative refinement and mitigating their error propagation. Moreover, the impacts of the number of PSPs and power allocation ratio on block error rate (BLER) performance are investigated to reveal the near-optimal pilot configuration. Simulation results show that the proposed PSP-SVT scheme outperforms existing full superimposed pilot-aided SVT schemes in terms of BLER with fast convergence speed.
\end{abstract}


\vspace{-0pt}
\section{Introduction}
Ultra-reliable and low-latency communication (URLLC) is a key service category for future wireless communication systems, supporting latency-sensitive applications such as autonomous driving and industrial control \cite{CYueWCM2023,DNguyen2022}. In these scenarios, transmitted data packets are typically very short, leaving limited room for pilot overhead and decoding latency. The problem is further aggravated in high-mobility environments, where rapid channel variation within a packet duration imposes additional challenges on reliable transmission.

Recently, the sparse vector transmission (SVT) \cite{Kim20201} has emerged as a promising technology for short-packet communications. In SVT, information bits are embedded into the indices of non-zero elements of a sparse vector and then spread by a random codebook, enabling reliable decoding through sparse signal recovery in the finite-blocklength regime. To improve transmission performance, several enhanced designs have been developed based on the SVT scheme. In \cite{Kim20202}, the enhanced SVT scheme incorporates modulated symbols into the non-zero entries to improve spectral efficiency. The sparse superposition coding (SSC) scheme proposed in \cite{ZhangXuewan2022} further enhances the reliability by increasing the minimum Euclidean distance between different non-zero elements, whereas the block orthogonal sparse superposition codes proposed in \cite{DhanTWC2023} exploit block orthogonality of codebook to improve the robustness of data decoding. Additional index dimensions have also been introduced to improve encoding efficiency, such as multi-mode index aided SVT scheme proposed in \cite{YangLinjie2024} and index redefined SSC scheme proposed in \cite{Zhangxue25}. Moreover, the block SVT scheme developed in \cite{yfzhang2025BSVC} maps information bits onto non-zero block indices rather than individual non-zero positions, thereby reducing the required codeword length and making it more reliable for short-packet transmission. Recently, SVT has also been extended to a variety of communication scenarios, including multiple-input multiple-output (MIMO) systems \cite{ZhangRuoyu2021}, grant-free access systems \cite{Luoyingzhe24}, and storage-constrained Internet of Things systems \cite{ZYFWCNC25}.

The aforementioned SVT schemes are primarily designed for quasi-static channels, where the channel remains approximately constant over one data packet. In high-mobility scenarios, however, the channel varies rapidly within the packet duration, making accurate channel acquisition more challenging and posing a key bottleneck for reliable decoding. To address this problem, we proposed a sparse superimposed vector transmission (SSVT) scheme in our prior \mbox{work \cite{ZhangYf2023},} where the rapidly time-varying channel is modeled by a basis expansion model (BEM) and full superimposed pilots (SP) are overlaid on all subcarriers to enable iterative channel estimation and data decoding. SSVT improves channel estimation performance, but it also introduces strong pilot–data interference over the entire bandwidth, which degrades the accuracy of data decoding. Recently, a joint pilot-data SVT (JPD-SVT) scheme proposed in \cite{JointPilotZhang2026} further improves the iterative receiver by selectively reusing a subset of reliably decoded data symbols as pseudo-pilots, thereby refining channel estimation and improving the block error rate (BLER) performance. Nevertheless, both SSVT and JPD-SVT schemes still rely on full SP, and the resulting strong pilot–data interference fundamentally limits the quality of the initial channel estimation and data decoding, which in turn constrains the subsequent iterative decoding performance.

Motivated by the above issues, in this paper, a partial superimposed pilot-aided SVT (PSP-SVT) scheme is proposed for short-packet URLLC over time-varying channels. Compared with SSVT \cite{ZhangYf2023} and JPD-SVT \cite{JointPilotZhang2026}, which employ full SPs over all subcarriers, the proposed PSP-SVT exploits the low-dimensional BEM representation and requires only a limited number of PSPs for channel estimation. This partial deployment reduces pilot–data interference and provides a more reliable initialization for subsequent iterative channel estimation and data decoding. The contributions of this paper are summarized as follows.
\begin{itemize}
\item A new PSP pattern is developed for high-mobility short-packet transmission. Unlike the full SP-aided SVT schemes in \cite{ZhangYf2023} and \cite{JointPilotZhang2026}, the proposed PSP pattern superimposes a small number of pilots over part of the subcarriers. This SP deployment method is well aligned with the low-dimensional BEM-based channel representation, where only a small number of channel coefficients need to be estimated. This effectively mitigates pilot–data interference, reduces the complexity of channel estimation, and provides a reliable initialization for the subsequent iterative decoding.

\item An iterative receiver is developed by leveraging the deployment characteristic of PSP. The reduced pilot–data interference enables more accurate initial BEM coefficient estimation, which in turn mitigates the error propagation in subsequent iterative channel estimation and data decoding. Furthermore, the impacts of the number of PSPs and power allocation ratio on BLER performance are investigated to reveal the near-optimal pilot configuration. Simulation results demonstrate that the proposed PSP-SVT scheme achieves superior BLER performance compared with existing SVT schemes.

\end{itemize}

The rest of this paper is organized as follows. In Section II, the encoding process of proposed PSP-SVT scheme is introduced. In Section III, the iterative joint channel estimation and data decoding algorithm for PSP-SVT is described. Extensive simulation results and discussions are shown in Section IV, while conclusions are drawn in Section V.

\vspace{-0pt}
\section{Encoding Process of PSP-SVT Scheme}

An orthogonal frequency division multiplexing (OFDM) system is considered, where each transmission block consists of $N_{\mathrm f}$ subcarriers. As shown in Fig.~1, a total of $b=(b_{\mathrm{I}}+b_{\mathrm{S}})$ information bits are mapped to a $K$-sparse data vector ${{\bf s}_{\mathrm d}}\in\mathbb{C}^{N}$, where $N_{\mathrm f}$ and $N$ denote the number of subcarriers and the sparse vector length, respectively. Let $\mathcal{D}$ denote the support set of the $K$ non-zero entries. The $b_{\mathrm{I}}$ index bits are conveyed by the active support pattern, satisfying $b_{\mathrm{I}}\le \left\lfloor \log_2 \binom{N}{K}\right\rfloor$, while the remaining $b_{\mathrm{S}}=K\log_2(M)$ bits are carried by the $K$ modulated symbols with modulation order $M$. The resulting sparse data vector is spread by a codebook ${\bf G}\in\mathbb{R}^{N_{\mathrm f}\times N}$, where ${\bf G}=[{\bf g}_1,\ldots,{\bf g}_N]$ and each codeword is composed of $1$ and $-1$ entries generated with equal probability. Accordingly, the coded data signal is given by ${\bf x}_{\mathrm{d}}={\bf G}{\bf s}_{\mathrm{d}} \in {\mathbb{C}^{{N_{\mathrm{f}}}}}$.

\begin{figure}[htbp]
	\centerline{\includegraphics[width=0.48\textwidth]{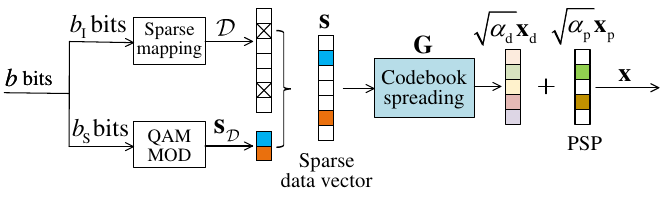}}
	\caption{Diagram of the encoding process of PSP-SVT scheme. }
	\label{fig1}
\vspace{-0pt}
\end{figure}

To track time-varying channels, $N_{\mathrm P}$ pilot symbols are superimposed on top of the coded data vector ${\bf x}_{\mathrm{d}}$.
Different from the existing full-SP SVT schemes in \cite{ZhangYf2023} and \cite{JointPilotZhang2026}, the proposed PSP-SVT scheme deploys pilots only on a subset of subcarriers. Specifically, the $N_{\mathrm P}$ PSPs are uniformly distributed over the $N_{\mathrm f}$ subcarriers to provide evenly distributed channel observations and avoid excessive overlap between the local pilot-induced regions. Nonuniform or clustered placement may lead to uneven channel observations and increased pilot–data interference. Let $\mathcal{P}=\{\mathcal{P}_1,\mathcal{P}_2,\ldots,\mathcal{P}_{N_{\mathrm P}}\}$ denote the index set of PSPs, where the spacing between adjacent non-zero pilot positions is $N_{\mathrm f}/N_{\mathrm P}$. The PSP vector ${\bf x}_{\mathrm{p}}\in\mathbb{C}^{N_{\mathrm f}}$ contains $N_{\mathrm P}$ non-zero pilot symbols only at the indices in $\mathcal{P}$, while the remaining entries are all zeros.

The transmitted frequency-domain signal is obtained by superimposing the coded sparse data vector and the PSP vector as ${\bf x}=\sqrt{\alpha_{\mathrm{d}}}{\bf x}_{\mathrm{d}}+\sqrt{\alpha_{\mathrm{p}}}{\bf x}_{\mathrm{p}}$, where $\alpha_{\mathrm{d}}$ and $\alpha_{\mathrm{p}}$ denote the data and pilot power allocation ratios, respectively, with $\alpha_{\mathrm{d}}+\alpha_{\mathrm{p}}=1$. By substituting ${\bf x}_{\mathrm{d}}={\bf G}{{\bf s}_{\mathrm d}}$, the transmitted signal can be equivalently expressed as
\begin{equation}
{\bf x}=\sqrt{\alpha_{\mathrm{d}}}\sum\nolimits_{k = 1}^{K}s_k{\bf g}_{k}+\sqrt{\alpha_{\mathrm{p}}}{\bf x}_{\mathrm{p}},
\label{eq_tx_sum}
\end{equation}
where $k\in\mathcal{D}$ denotes the $k$-th non-zero index.

After inverse discrete Fourier transform (IDFT), the transmitted time-domain signal is given by ${\bf x}_{\mathrm{T}}={\bf F}^{\mathrm H}{\bf x}$, where ${\bf F}\in\mathbb{C}^{N_{\mathrm f}\times N_{\mathrm f}}$ is a normalized DFT matrix. After cyclic prefix (CP) insertion, the signal ${\bf x}_{\mathrm{T}}$ is transmitted through the time-varying channel. Following CP removal and DFT, the received frequency-domain signal can be written as
\begin{equation}
{\bf{y}} = {\bf{F}}{{\bf{H}}_{\rm{T}}}{{\bf{F}}^{\mathrm H}}{\bf{x}} + {\bf{z}} = {{\bf{H}}_{\rm{F}}}{\bf{x}} + {\bf{z}},
\label{eq2}
\end{equation}
where ${{\bf{H}}_{\rm{T}}} = \sum\nolimits_{l = 0}^{L - 1} {{{\bf{\Pi }}^l}{\rm{diag}}({{\bf{h}}_l})}$ is the time-domain channel matrix with $L$ taps, ${{{\bf{h}}_l}}$ is the channel impulse response (CIR) of the $l$-th tap, ${\bf{\Pi }} = {\mathop{\rm circ}\nolimits} \{ [0,1,0, \cdots ,0]_{{N_{\mathrm f}} \times 1}^T\} $ is a permutation matrix,  ${{\bf{H}}_{\rm{F}}} = {\bf{F}}{{\bf{H}}_{\rm{T}}}{{\bf{F}}^{\mathrm H}} \in {\mathbb{C}^{{N_{\mathrm f}} \times {N_{\mathrm f}}}}$ is the frequency-domain channel matrix, and ${\bf{z}} \sim {\cal C}{\cal N}(0,{\sigma ^2}{{\bf{I}}_{N_{\mathrm f}}})$ is the additive white Gaussian noise (AWGN) vector with variance $\sigma ^2$, and $\mathbf I_{N_{\mathrm f}}$ denotes the ${N_{\mathrm f}}\times {N_{\mathrm f}}$ identity matrix.

Note that the number of time-varying channel coefficients $N_{\mathrm f}L$ is much larger than the observation dimension $N_{\mathrm f}$, making channel estimation difficult.
To reduce the number of unknown channel coefficients, the CIR of each path is modeled using the Slepian BEM \cite{ZYFTWC25}. 
This choice is motivated by its optimal energy concentration property for band-limited Doppler processes, which leads to faster eigenvalue decay and a more compact representation than conventional BEMs \cite{ZYFTWC25}. Accordingly, the CIR can be expressed as
\begin{equation}
{{\bf{h}}_l} = \sum\nolimits_{q = 0}^{Q - 1} {{c_{q,l}}{{\bf{b}}_q}}  + {{\bf{w}}_l},
\end{equation}
where $Q$ ($Q\ll N_{\mathrm f}$) is the BEM order, $c_{q,l}$ denotes the $q$-th Slepian coefficient of the $l$-th path, ${\bf b}_q$ is the $q$-th Slepian basis function, and ${\bf w}_l$ represents the modeling error. The Slepian basis ${\bf b}_q$ is obtained from the eigenvalue problem
\begin{equation}
\sum_{n=0}^{N_{\mathrm f}-1}\frac{\sin\!\left(2\pi f_{\max}(n-m)\right)}{\pi(n-m)}b_q[n]
=\lambda_q b_q[m],
\label{eq_slepian}
\end{equation}
where $f_{\max}$ is the maximum Doppler frequency and $\lambda_q$ is the corresponding eigenvalue of ${\bf b}_q$.

Based on the Slepian-BEM modeling, the time-domain channel matrix ${\bf H}_{\mathrm T}$ in (2) can be further expressed as
\begin{equation}
\mathbf H_{\mathrm T}
=
\sum\nolimits_{q = 0}^{Q - 1}
\operatorname{diag}(\mathbf b_q)\mathbf F^{\mathrm H}
\operatorname{diag}(\mathbf F_L\mathbf c_q)\mathbf F
+\mathbf W,
\end{equation}
where ${\bf F}_L$ is a submatrix composed of the first $L$ columns of ${\bf F}$, ${\bf c}_q=[c_{q,0},\ldots,c_{q,L-1}]^{\mathrm T}$, and ${\bf W}$ denotes the modeling error matrix. Substituting (5) into (2), the received signal can be rewritten as
\begin{equation}
{\bf y}
=
\sum\nolimits_{q = 0}^{Q - 1}
{\bf \bar B}_q
\mathrm{diag}\!\left(
\sqrt{\alpha_{\mathrm d}}{\bf x}_{\mathrm d}
+
\sqrt{\alpha_{\mathrm p}}{\bf x}_{\mathrm p}
\right)
{\bf F}_L{\bf c}_q
+
{\bf u},
\label{eq_bem_rx}
\end{equation}
where ${\bf \bar B}_q={\bf F}\mathrm{diag}({\bf b}_q){\bf F}^H$, and ${\bf u}$ includes both the BEM modeling error and AWGN.

Thanks to the Slepian-BEM based channel modeling, the number of unknown channel parameters is reduced from $N_{\mathrm f}L$ to $QL$. In finite-scattering environments, $QL$ is typically much smaller than $N_{\mathrm f}$. Therefore, only $N_{\mathrm P}\ge QL$ pilots are sufficient to estimate all the BEM coefficients. Based on the signal model in (6), an iterative joint channel estimation and data decoding algorithm is developed to achieve reliable decoding for PSP-SVT.

\section{Iterative Joint Channel Estimation and Data Decoding}
In this section, an iterative joint channel estimation and data decoding receiver is proposed for PSP-SVT. The receiver operates in two stages. At the initial stage, a coarse channel estimate is obtained by exploiting the received PSP signal, followed by initial data decoding based on the estimated channel. At the iterative stage, the decoded data symbols are used as pseudo-pilots to further refine the channel estimation and improve the subsequent data decoding.

\vspace{-0pt}
\subsection{Initial Channel Estimation and Data Decoding}
At the receiver, the $m$-th received PSP signal can be expressed as
\begin{equation}
y_{\mathrm p}(m)
=
\sqrt{\alpha_{\mathrm p}}x_{\mathrm p}(n)
\sum\nolimits_{q = 0}^{Q - 1}
{\bar h}_{q,n}\,
\beta_q\!\left(\bmod(m-n, N_{\mathrm f})\right),
\label{eq_yp_scalar}
\end{equation}
where $\mathrm{mod}(a,b)$ returns the remainder after dividing $a$ by $b$, $\beta_q(\bmod(m-n, N_{\mathrm f}))$ denotes the $(m,n)$-th entry of the equivalent BEM spreading matrix ${\bf \bar B}_q$, which depends on the circular index difference $\bmod(m-n, N_{\mathrm f})$, and ${\bar h}_{q,n}$ denotes the equivalent frequency-domain channel coefficient of the $q$-th Slepian-BEM basis corresponding to the input PSP position $\mathcal{P}_n$, i.e.,
\begin{equation}
{\bar h}_{q,n}
=
\sum\nolimits_{l = 0}^{L - 1}
c_{q,l}e^{-j\frac{2\pi \mathcal{P}_n l}{N_{\mathrm f}}}
\label{eq_hqn}
\end{equation}

\begin{figure}[htbp]
	\centerline{\includegraphics[width=0.49\textwidth]{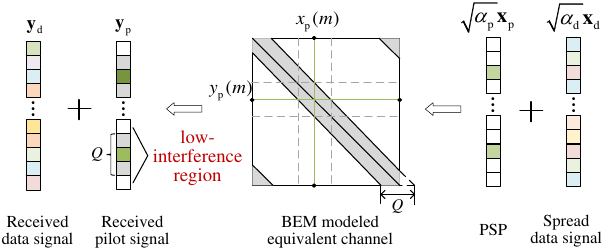}\vspace{-5pt}}
	\caption{Schematic of the PSP and spread data signal transmission process. }
\vspace{-0pt}
\end{figure}

The transmitted PSP symbols and the spread data signal are jointly distorted by the Slepian-BEM modeled channel matrix, as illustrated in Fig.~2. Owing to the energy concentration property of the Slepian-BEM, the equivalent frequency-domain channel matrix is approximately banded, with its effective bandwidth determined by the BEM order $Q$. Accordingly, the dominant energy of each PSP is concentrated within $Q$ neighboring subcarriers around its aligned output position. On the remaining subcarriers, the PSP-induced components are nearly zero. Consequently, the interference introduced by the PSP symbols to the received data signal is very weak outside these local regions. These regions are referred to as \emph{low-interference regions}, as highlighted in Fig.~2. Such low-interference regions enable more accurate initial BEM coefficient estimation and reliable data decoding, which in turn provides a reliable starting point for the subsequent iterative decoding process and effectively mitigates the error propagation between channel estimation and data decoding.

Accordingly, the receiver collects the $Q$ neighboring received signals associated with each PSP position to construct the initial channel estimation model. Specifically, let $\mathcal{S}$ denote the union of the local received regions associated with all PSP positions, i.e.,
\begin{equation}
\mathcal{S}
=
\bigcup_{n=1}^{N_{\mathrm P}}
\left\{
\mathrm{mod}(\mathcal{P}_n+q,N_{\mathrm f})
\mid
q=-\left\lfloor\frac{Q-1}{2}\right\rfloor,\ldots,\left\lfloor\frac{Q}{2}\right\rfloor
\right\}.
\label{eq_S_set}
\end{equation}

By extracting the received signal indexed by $\mathcal{S}$, the initial channel estimation model is formulated as
\begin{equation}
{\bf \bar{y}}_{\mathrm p}
=
{\bf A}^{(0)}{\bf c}
+ {\bf \bar{y}}_{\mathrm d}
+
{\bf u},
\label{eq_yAc_model}
\end{equation}
where ${{{\bf{\bar y}}}_{\rm{p}}} = {{\bf{P}}_{\cal S}}{{\bf{y}}_{\rm{p}}}$, ${{{\bf{\bar y}}}_{\rm{d}}} = {{\bf{P}}_{\cal S}}{{\bf{y}}_{\rm{d}}}$ is the received data signal, ${{\bf{P}}_{\cal S}}\in{\mathbb{R}}^{{|\cal S|}\times N_{\mathrm f}}$ is a row selection matrix determined by the index set ${\cal S}$, ${\bf{c}} = {[{\bf{c}}_0^{\rm{H}}, \cdots ,{\bf{c}}_{Q - 1}^{\rm{H}}]^{\rm{H}}} \in {{\mathbb C}^{QL}}$, and ${{\bf{A}}^{(0)}} = [{{\bf{A}}_0^{(0)}},{{\bf{A}}_1^{(0)}}, \cdots ,{{\bf{A}}_{Q - 1}^{(0)}}] \in {{\mathbb{C}}^{{{|\cal S|}} \times QL}}$ is the measurement matrix. The $q$-th submatrix of $\bf A$ is given by
\begin{equation}
{{\bf{A}}_q^{(0)}} = {{\bf{P}}_{\cal S}}{\bf{F}}{\mathop{\rm diag}\nolimits} ({{\bf{b}}_q}){{\bf{F}}^{\rm{H}}}{\rm{diag}}(\sqrt {{\alpha _{\rm{p}}}} {{\bf{x}}_{\rm{p}}}){{\bf{F}}_L}
\end{equation}
for $q = 0,1, \cdots ,Q - 1$. Based on the signal model in (10), the initial BEM coefficient estimate at the $0$-th iteration is obtained by
\begin{equation}
{{{\bf{\hat c}}}^{(0)}} = {\left( {{{({{\bf{A}}^{(0)}})}^{\rm{H}}}{{\bf{A}}^{(0)}}} \right)^{ - 1}}{({{\bf{A}}^{(0)}})^{\rm{H}}}{{{\bf{\bar y}}}_{\rm{p}}}.
\end{equation}
With $\hat{\bf c}^{(0)}$ in hand, the initial frequency-domain channel matrix can be reconstructed as
\begin{equation}
{\bf{\hat H}}_{\rm{F}}^{(0)} = \sum\nolimits_{q = 0}^{Q - 1} {{{{\bf{\bar B}}}_q}{\rm{diag}}({{\bf{F}}_L}{\bf{\hat c}}_q^{(0)})},
\end{equation}
where $\hat{\mathbf c}^{(0)}_q$ denotes the $q$-th subvector of $\hat{\mathbf c}^{(0)}$.

The received data signal is then obtained by removing the received pilot signal from $\bf y$, i.e.,
\begin{equation}
{\bf{\hat y}}_{\rm{d}}^{(0)} = {\bf{y}} - \sqrt {{\alpha _{\rm{p}}}} {\bf{\hat H}}_{\rm{F}}^{(0)}{{\bf{x}}_{\rm{p}}}.
\end{equation}

The initial data decoding is performed based on the reconstructed channel matrix ${\bf{\hat H}}_{\rm F}^{(0)}$. Specifically, the data sparse vector is recovered by solving
\begin{equation}
\{ {\bf{\hat s}}_{\rm{d}}^{(0)},{{\hat {\cal D}}^{(0)}}\}  = \mathop {\arg \min }\limits_{||{\bf{s}}_{\rm{d}}^{(0)}|{|_0} = K} \left\| {{\bf{\hat y}}_{\rm{d}}^{(0)} - \sqrt {{\alpha _{\rm{d}}}} {\bf{\hat H}}_{\rm{F}}^{(0)}{\bf{Gs}}_{\rm{d}}^{(0)}} \right\|_2^2.
\end{equation}
In this paper, the multipath matching pursuit (MMP) algorithm \cite{Kim20202} is used to solve the problem in (15). Based on the detected support set $\hat{\mathcal D}^{(0)}$, the corresponding $k$-th QAM symbol is demodulated as
\begin{equation}
\tilde s_{{\mathrm d},k}^{(0)}
=
\arg\min_{a\in\mathcal A}
\left|
\hat s_{{\mathrm d},k}^{(0)}-a
\right|^2,
\quad
k=1,2,\ldots,K,
\label{eq_qam_demod}
\end{equation}
where $\mathcal A$ denotes the $M$-ary QAM constellation alphabet.

\vspace{-0pt}
\subsection{Iterative Enhanced Data Decoding}

For the $j$-th iteration ($j>0$), the decoded sparse data vector obtained from the previous iteration is further exploited as pseudo-pilots to refine the channel estimation. Specifically, the reconstructed spread data signal from the $(j-1)$-th iteration is obtained as
\begin{equation}
{\bf{\hat x}}_{\rm d}^{(j-1)}
=
{\bf G}{\bf{\hat s}}_{\rm d}^{(j-1)}.
\end{equation}
By combining the original PSP and the reconstructed spread data signal, a new pilot signal is formed as
\begin{equation}
{\bf x}_{\rm ep}^{(j)}
=
\sqrt{\alpha_{\rm p}}{\bf x}_{\rm p}
+
\sqrt{\alpha_{\rm d}}{\bf{\hat x}}_{\rm d}^{(j-1)}.
\label{eq_xep}
\end{equation}

Based on the new pilot signal ${\bf x}_{\rm ep}^{(j)}$, the received signal model for the $j$-th channel estimation is formulated as
\begin{equation}
{\bf{ y}}
=
{\bf A}^{(j)}{\bf c}
+
{\bf u},
\label{eq_iter_model}
\end{equation}
where the new measurement matrix is written as ${\bf A}^{(j)}=[{\bf A}_0^{(j)},{\bf A}_1^{(j)},\ldots,{\bf A}_{Q-1}^{(j)}],$ with its $q$-th submatrix given by
\begin{equation}
{\bf A}_q^{(j)}
=
{\bf F}
\mathrm{diag}({\bf b}_q)
{\bf F}^{\rm H}
\mathrm{diag}\!\left({\bf x}_{\rm ep}^{(j)}\right)
{\bf F}_L,
\end{equation}
for $q=0,1,\ldots,Q-1$.
Accordingly, the BEM coefficient vector is updated as
${\bf{\hat c}}^{(j)}=\left(({\bf A}^{(j)})^{\rm H}{\bf A}^{(j)}\right)^{-1}({\bf A}^{(j)})^{\rm H}{\bf{ y}}$. 
Using ${\bf{\hat c}}^{(j)}$, the channel matrix ${\bf{\hat H}}_{\rm F}^{(j)}$ can be reconstructed.

The data decoding is then performed in a similar manner to (14)--(16) based on the updated channel matrix ${\bf{\hat H}}_{\rm F}^{(j)}$. The above channel estimation and data decoding are iteratively executed until the maximum iteration number $N_{\mathrm{iter}}$ is reached or the convergence criterion
\begin{equation}
{{||{{{\bf{\hat c}}}^{(j + 1)}} - {{{\bf{\hat c}}}^{(j)}}||_2^2} \mathord{\left/
 {\vphantom {{||{{{\bf{\hat c}}}^{(j + 1)}} - {{{\bf{\hat c}}}^{(j)}}||_2^2} {||{{{\bf{\hat c}}}^{(j)}}||_2^2}}} \right.
 \kern-\nulldelimiterspace} {||{{{\bf{\hat c}}}^{(j)}}||_2^2}} \le \epsilon
\end{equation}
is satisfied, where $\epsilon$ is a predefined threshold, e.g., $\epsilon=10^{-4}$.

\textit{Remark:}
The proposed PSP pattern concentrates the dominant pilot energy within $Q$ neighboring subcarriers around each PSP, creating low-interference regions for initial channel estimation and data decoding. This enables the decoded data symbols to serve as effective pseudo-pilots in subsequent iterations, thereby suppressing the error propagation between channel estimation and data decoding. Consequently, the proposed PSP-SVT scheme achieves faster convergence and improved robustness over existing full-SP SVT schemes as verified by the simulation results in Section IV.

\vspace{-0pt}
\section{Simulation Results}
In this section, Monte Carlo simulations are conducted to evaluate the performance of the proposed PSP-SVT scheme, including the BLER, convergence behavior, and the optimal number of PSPs and power allocation ratio. The existing pilot-less SSC scheme \cite{ZhangXuewan2022}, SSVT \cite{ZhangYf2023}, and JPD-SVT \cite{JointPilotZhang2026} are selected as baseline schemes. The simulation parameters are specified as follows. The carrier frequency is 4.9 GHz, and the subcarrier spacing is 15 kHz. The sparse vector length is fixed at $N=132$, the user velocity is set to $v=300$ km/h, the BEM order is set to $Q=2$, and the $K$ non-zero elements are modulated using QPSK. The simulations are conducted over the 5G TDL-B channel \cite{5GTDL}, which consists of $L=6$ paths and follows the Jakes' Doppler spectrum.

Fig.~3 compares the proposed PSP-SVT scheme with the SSVT scheme \cite{ZhangYf2023} and JPD-SVT scheme \cite{JointPilotZhang2026}, where the baseline schemes use $84$ SPs while PSP-SVT scheme employs $14$ PSPs under the same sparsity level $K=2$. Despite using far fewer SPs, PSP-SVT achieves better BLER performance at different SNR values. 
It is observed that the BLER first decreases and then increases with  $\alpha_{\mathrm d}$. A small $\alpha_{\mathrm d}$ leads to a low effective data SNR, whereas an excessively large $\alpha_{\mathrm d}$ reduces the pilot power and degrades channel estimation accuracy. The optimal data power allocation ratio of PSP-SVT is higher than that of SSVT, i.e., about $0.7$ versus $0.63$. This indicates that the proposed PSP-SVT scheme allows more power to be allocated to data transmission while still maintaining reliable channel estimation. In subsequent simulations, $\alpha_{\mathrm d}=0.7$ is adopted for PSP-SVT scheme. 

Fig.~4 shows the impact of the number of PSP $N_{\mathrm P}$ on BLER performance. When $N_{\mathrm P}<6$, the BLER performance degrades sharply because the number of pilot observations is insufficient for accurate BEM coefficient estimation, i.e., $N_{\mathrm P}<QL$. Once sufficient pilot observations are available for BEM coefficient estimation, further increasing $N_{\rm P}$ provides limited gain but increases pilot-data interference and shrinks the low-interference regions, thereby degrading the BLER performance. These results indicate that the number of PSPs and power allocation ratio should be jointly optimized in PSP-SVT system. Accordingly, $N_{\mathrm P}=14$ is adopted in the subsequent simulations, as it provides a favorable BLER tradeoff across different SNRs.

\begin{figure}[htbp]
	\centerline{\includegraphics[width=0.48\textwidth]{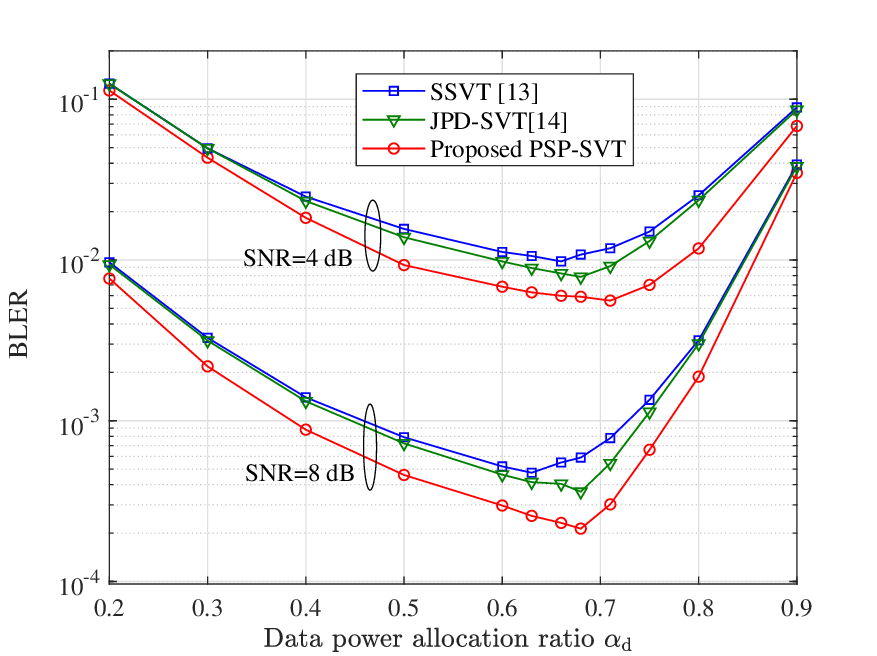}}
	\caption{BLER versus $\alpha_{\rm d}$ with $N_{\mathrm P}=14$. \vspace{-0pt}}
	\label{fig3}
\vspace{-0pt}
\end{figure}

\begin{figure}[htbp]
	\centerline{\includegraphics[width=0.48\textwidth]{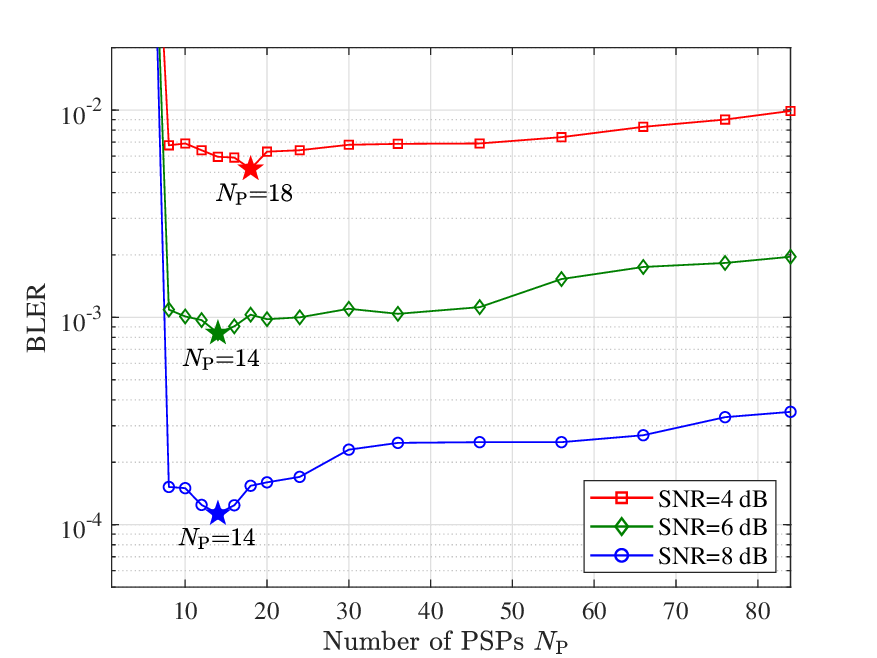}}
	\caption{BLER as a function of the number of PSPs with $\alpha_{\mathrm d}=0.7$. \vspace{-0pt}}
	\label{fig4}
\vspace{-0pt}
\end{figure}

Fig.~5 compares the BLER performance of different schemes. The pilot-less SSC scheme \cite{ZhangXuewan2022} performs poorly because no channel estimation is involved, making it ineffective over rapidly time-varying channels. In contrast, all pilot-aided schemes can maintain reliable decoding by tracking the channel variation. Compared with SSVT \cite{ZhangYf2023} and JPD-SVT \cite{JointPilotZhang2026}, the proposed PSP-SVT achieves the best BLER performance at different SNR values. This gain mainly comes from the PSP pattern, which reduces the pilot-data interference by deploying a small number of SPs. As a result, the iterative receiver starts from a more accurate channel estimation, which effectively suppresses the error propagation between channel estimation and data decoding. This indicates that the number of SPs is critical for balancing the channel estimation accuracy and decoding reliability.

Fig.~6 compares the BLER performance for different numbers of subcarriers. The proposed PSP-SVT outperforms the JPD-SVT scheme for all tested $N_{\mathrm f}$ values. Moreover, the performance gain becomes increasingly pronounced as $N_{\mathrm f}$ grows. This is because, with a fixed number of SPs, the pilot deployment in PSP-SVT becomes sparser at larger $N_{\mathrm f}$, which enlarges the low-interference regions between pilots and data. As a result, the initialization quality of both channel estimation and data decoding is further improved, making the advantage of the proposed PSP design more pronounced with a larger number of subcarriers.

\begin{figure}[htbp]
	\centerline{\includegraphics[width=0.48\textwidth]{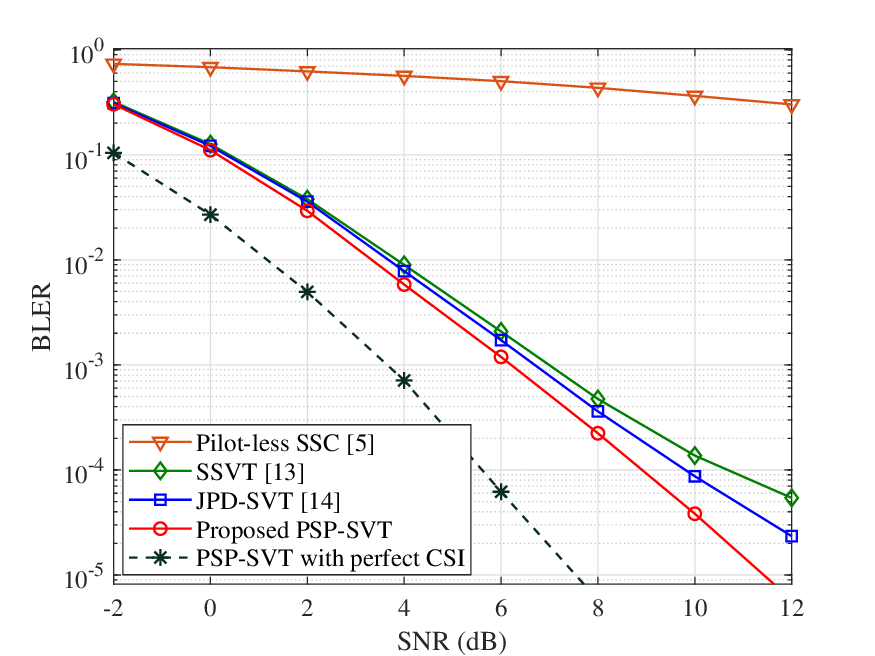}}
	\caption{BLER versus SNR with $K=2$, $N_{\mathrm P}=14$, $N_{\mathrm f}=84$ and $b=17$. \vspace{-5pt}}
	\label{fig5}
\vspace{-0pt}
\end{figure}

\begin{figure}[htbp]
	\centerline{\includegraphics[width=0.48\textwidth]{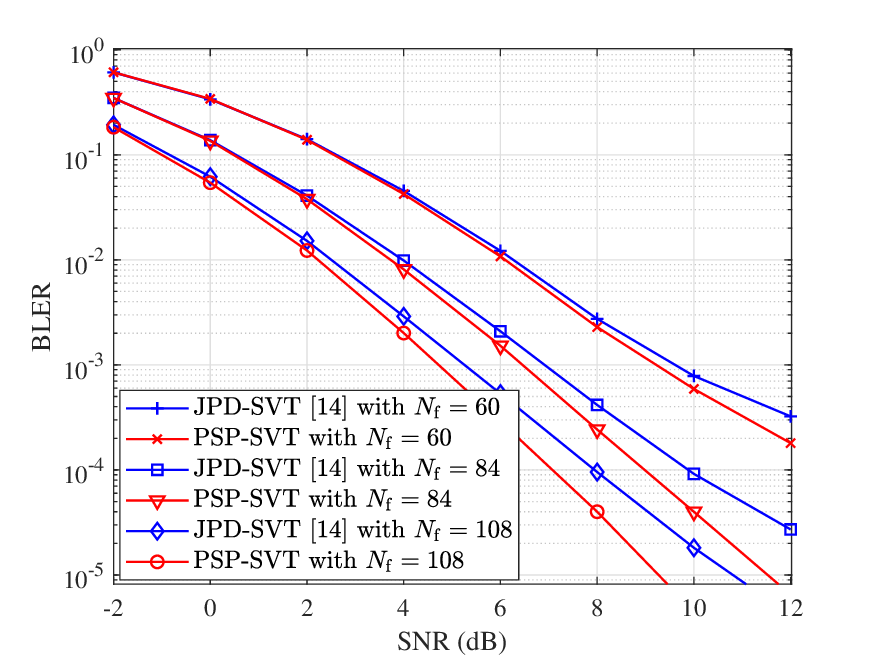}}
	\caption{BLER versus SNR under different $N_{\mathrm f}$ with $K=2$ and $b=17$. }
	\label{fig5}
\vspace{-0pt}
\end{figure}

Fig.~7 compares the convergence speed of different schemes. It can be observed that the proposed PSP-SVT achieves a lower initial BLER at the $0$-th iteration than the SSVT scheme for both $K=2$ and $K=3$. This confirms that the PSP pattern provides a more accurate initial channel estimation and data decoding by reducing the pilot--data interference in the received signal. Note that both schemes converge very fast and typically reach stable performance within two iterations, while the proposed PSP-SVT scheme converges to a lower BLER level. It is noteworthy that the receiver burden mainly arises from BEM-based channel estimation and sparse data decoding in each iteration. Since PSP-SVT typically converges within two iterations, its iterative processing burden remains limited. For other mobility conditions, the same receiver structure remains applicable, while the BEM order and the number of PSPs can be adjusted according to the channel variation rate.

\begin{figure}[htbp]
	\centerline{\includegraphics[width=0.48\textwidth]{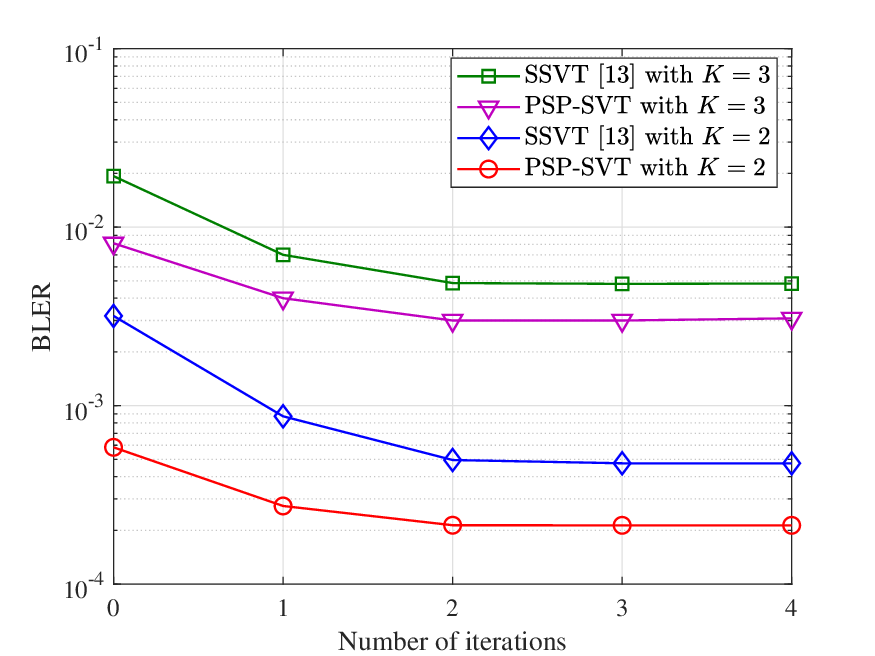}}
	\caption{Convergence speed of different schemes under different $K$. }
	\label{fig6}
\vspace{-0pt}
\end{figure}

\vspace{-0pt}
\section{Conclusions}

In this paper, a PSP-SVT scheme is proposed for short-packet URLLC over high-mobility channels. By sparsely deploying a small number of pilots over a subset of subcarriers, the proposed PSP pattern effectively reduced pilot--data interference and created low-interference regions for receiver initialization. Based on this structure, an iterative joint channel estimation and data decoding receiver is developed under the Slepian-BEM channel modeling, where the decoded data symbols are further reused as pseudo-pilots for iterative refinement. Simulation results show that the proposed PSP-SVT scheme outperforms existing full-SP aided SVT schemes in terms of BLER performance with fast convergence speed.

\vspace{-0pt}
\bibliographystyle{IEEEbib}
\bibliography{refs}

@ARTICLE{JointPilotZhang2026,
  author={X. Zhang and K. Wang},
  journal={IEEE Commun. Lett.}, 
  title={A Joint Pilot-Data Sparse Vector Transmission Framework for Short-Packet URLLC}, 
  year={Mar. 2026},
  volume={30},
  number={},
  pages={1265-1269},
  doi={10.1109/LCOMM.2026.3670309}}

@ARTICLE{ZYFTWC25,
  author={Y. Zhang and X. Zhu and Y. Liu and Y. Guan and G. David and V. K. N. Lau},
  journal={IEEE Trans. Wireless Commun.}, 
  title={Basis Expansion Extrapolation based Long-Term Channel Prediction for Massive {MIMO OTFS} Systems}, 
  year={Feb. 2026},
  volume={25},
  number={},
  pages={2280-2296},
  doi={10.1109/TWC.2025.3595756}}

@ARTICLE{ZhangYf2023,
  author={Zhang, Y. and Zhu, X. and Liu, Y. and Jiang, Y. and Guan, Y. and González G., D. and Lau, V. K. N.},
  journal={IEEE Wireless Commun. Lett.}, 
  title={Sparse Superimposed Vector Transmission for Short-Packet High-Mobility Communication}, 
  year={Nov. 2023},
  volume={12},
  number={11},
  pages={1961-1965},
  doi={10.1109/LWC.2023.3303234}}

@INPROCEEDINGS{ZYFWCNC25,
  author={Y. Zhang and X. Fan and H. Liang and W. Yang and J. Zheng and T. H. Luan},
  booktitle={2025 IEEE Wireless Communications and Networking Conference (WCNC)}, 
  title={Semi-Tensor Sparse Vector Coding for Short-Packet URLLC with Low Storage Overhead}, 
  year={Milan, Italy, 2025},
  volume={},
  number={},
  pages={1-6},
  doi={10.1109/WCNC61545.2025.10978477}}

@INPROCEEDINGS{Luoyingzhe24,
  author={Y. Luo and X. Zhu and Y. Zhang and Z. Guo},
  booktitle={IEEE International Conference on Communications (ICC)}, 
  title={Sparse Vector Coding Based Massive Grant-Free Access for Short-Packet Communication in {IIoT}}, 
  year={Denver, CO, USA, 2024},
  volume={},
  number={},
  pages={5395-5400},
  doi={10.1109/ICC51166.2024.10622836}}

@ARTICLE{yfzhang2025BSVC,
  author={Y. Zhang and X. Zhu and Y. Liu and X. Fan and Y. Guan and M. L. Wong and V. K. N. Lau},
  journal={IEEE Trans. Commun.}, 
  title={Block Sparse Vector Codes for Ultra-Reliable and Low-Latency Short-Packet Transmission}, 
  year={Oct. 2025},
  volume={73},
  number={10},
  pages={9750-9766},
  doi={10.1109/TCOMM.2025.3562520}}

@ARTICLE{Zhangxue25,
  author={X. Zhang and C. Yue and J. Guo and M. Shirvanimoghaddam and Y. Li},
  journal={IEEE Trans. Commun.}, 
  title={Generalized Index Redefinition-Based Sparse Mapping for Sparse Vector Transmission}, 
  year={Aug. 2025},
  volume={73},
  number={8},
  pages={5920-5934},
  doi={10.1109/TCOMM.2025.3541056}}

@Article{5GTDL,
  author={},
  journal={Standard 3GPP TR 38.901}, 
  title={{Study on Channel Model for Frequencies From 0.5 to 100 {GHz}}}, 
  year={2017.},
  volume={},
  number={},
  pages={},
 }

@ARTICLE{CYueWCM2023,
  author={C. Yue and V. Miloslavskaya and M. Shirvanimoghaddam and B. Vucetic and Y. Li},
  journal={IEEE Commun. Mag.}, 
  title={Efficient Decoders for Short Block Length Codes in {6G} {URLLC}}, 
  year={Apr. 2023},
  volume={61},
  number={4},
  pages={84-90},
  doi={10.1109/MCOM.001.2200275}}

@ARTICLE{DNguyen2022,
  author={D. Nguyen and M. Ding and P. Pathirana and A. Seneviratne and J. Li and D. Niyato and O. Dobre and H. Poor},
  journal={IEEE Internet Things J.}, 
  title={{6G} Internet of Things: A Comprehensive Survey}, 
  year={Jan. 2022},
  volume={9},
  number={1},
  pages={359-383},
  doi={10.1109/JIOT.2021.3103320}}

@ARTICLE{DhanTWC2023,
  author={D. Han and J. Park and Y. Lee and H. Poor and N. Lee},
  journal={IEEE Trans. Commun.}, 
  title={Block Orthogonal Sparse Superposition Codes for Ultra-Reliable Low-Latency Communications}, 
  year={Dec. 2023},
  volume={71},
  number={12},
  pages={6884-6897},
  doi={10.1109/TCOMM.2023.3317912}}

@ARTICLE{Kim20201,
  author={W. Kim and H. Ji and H. Lee and Y. Kim and J. Lee and B. Shim},
  journal={IEEE Veh. Technol. Mag.}, 
  title={Sparse Vector Transmission: An Idea Whose Time Has Come}, 
  year={Sep. 2020},
  volume={15},
  number={3},
  pages={32-39},
  doi={10.1109/MVT.2020.2976891}}

@ARTICLE{ZhangXuewan2022,
  author={X. Zhang and D. Zhang and B. Shim and G. Han and D. Zhang and T. Sato},
  journal={IEEE Internet Things J.}, 
  title={Sparse Superimposed Coding for Short-Packet {URLLC}}, 
  year={Apr. 2022},
  volume={9},
  number={7},
  pages={5275-5289},
  doi={10.1109/JIOT.2021.3108161}}

@ARTICLE{YangLinjie2024,
  author={Yang, L. and Fan, P.},
  journal={IEEE Commun. Lett.}, 
  title={Multiple-Mode Sparse Superposed Code With Low Block Error Rate for Short Packet {URLLC}}, 
  year={Feb. 2024},
  volume={28},
  number={2},
  pages={248-252},
  doi={10.1109/LCOMM.2023.3348570}}

@ARTICLE{Kim20202,
  author={W. Kim and S. Bandari and B. Shim},
  journal={IEEE Trans. Veh. Technol.}, 
  title={Enhanced Sparse Vector Coding for Ultra-Reliable and Low Latency Communications}, 
  year={May 2020},
  volume={69},
  number={5},
  pages={5698-5702},
  doi={10.1109/TVT.2020.2982943}}

@ARTICLE{ZhangRuoyu2021,
  author={R. Zhang and B. Shim and Y. Lou and S. Jia and W. Wu},
  journal={IEEE Trans. Veh. Technol.}, 
  title={Sparse Vector Coding Aided Ultra-Reliable and Low-Latency Communications in Multi-User Massive {MIMO} Systems}, 
  year={Jan. 2021},
  volume={70},
  number={1},
  pages={1019-1024},
  doi={10.1109/TVT.2020.3044190}}

@ARTICLE{l,
  author={X. Zhang and D. Zhang},
  journal={IEEE Commun. Lett.}, 
  title={Sparse Superimposed Coding Based on Index Redefinition}, 
  year={2023},
  volume={},
  number={},
  pages={1-1},
  doi={10.1109/LCOMM.2023.3257314}}

\end{document}